\documentclass[%
 aip,
 sd,
 amsmath,amssymb,
 reprint,%
 numerical,
]{revtex4-1}

\usepackage{graphicx,}
\usepackage{dcolumn}
\usepackage{bm}
\usepackage{float}
\usepackage{amssymb}
\usepackage{mathrsfs}
\usepackage[usenames, dvipsnames]{color}

\begin{document}

\preprint{AIP/123-QED}
\title{Exciton Delocalization Incorporated Drift-Diffusion Model for Bulk-Heterojunction Organic Solar Cells} 
\author{Zi Shuai Wang}%
\affiliation{Department of Electrical and Electronic Engineering, The University of Hong Kong, Pokfulam Road, Hong Kong}
\author {Wei E. I. Sha}%
\email{wsha@eee.hku.hk}
\affiliation{Department of Electrical and Electronic Engineering, The University of Hong Kong, Pokfulam Road, Hong Kong}
\author {Wallace C. H. Choy}%
\email{chchoy@eee.hku.hk}
\affiliation{Department of Electrical and Electronic Engineering, The University of Hong Kong, Pokfulam Road, Hong Kong}
\begin{abstract}
Modeling the charge-generation process is highly important to understand device physics and optimize power conversion efficiency of bulk-heterojunction (BHJ) organic solar cells (OSCs). Free carriers are generated by both ultrafast exciton delocalization and slow exciton diffusion and dissociation at the heterojunction interface. In this work, we developed a systematic numerical simulation to describe the charge-generation process by a modified drift-diffusion model. The transport, recombination, and collection of free carriers are incorporated to fully capture the device response. The theoretical results match well with the state-of-the-art high-performance organic solar cells. It is demonstrated that the increase of exciton delocalization ratio reduces the energy loss in the exciton diffusion-dissociation process, and thus, significantly improves the device efficiency especially for the short-circuit current. By changing the exciton delocalization ratio, OSC performances are comprehensively investigated under the conditions of short-circuit and open-circuit. Particularly, bulk recombination dependent fill factor saturation is unveiled and understood. As a fundamental electrical analysis of the delocalization mechanism, our work is important to understand and optimize the high-performance OSCs.
\end{abstract}

\pacs{}

\maketitle 

\section{Introduction}
Due to the advantages of low cost, easy fabrication, mechanical flexibility, organic solar cells have attracted a lot of attention in the last decades. Among countless organic materials, the most successful ones are the conjugated polymers, which were widely adopted in bulk heterojunction in conjugation with the fullerene \cite{Heeger1}. BHJ structure greatly facilitates the dissociation of photogenerated excitons when they diffuse to the interface before decaying to the ground state. Different from the exciton diffusion-dissociation process, the exciton delocalization in an ultrafast timescale has been observed recently \cite{Zhong5,Askat,Kaake6,Kaake7,Kaake8,Kaake9,Bakulin9,FriendNew}. The delocalized excitons have a great influence on the charge separation and device performance of BHJ organic solar cells.

\begin{figure}[!hbp]
\includegraphics[width=6cm]{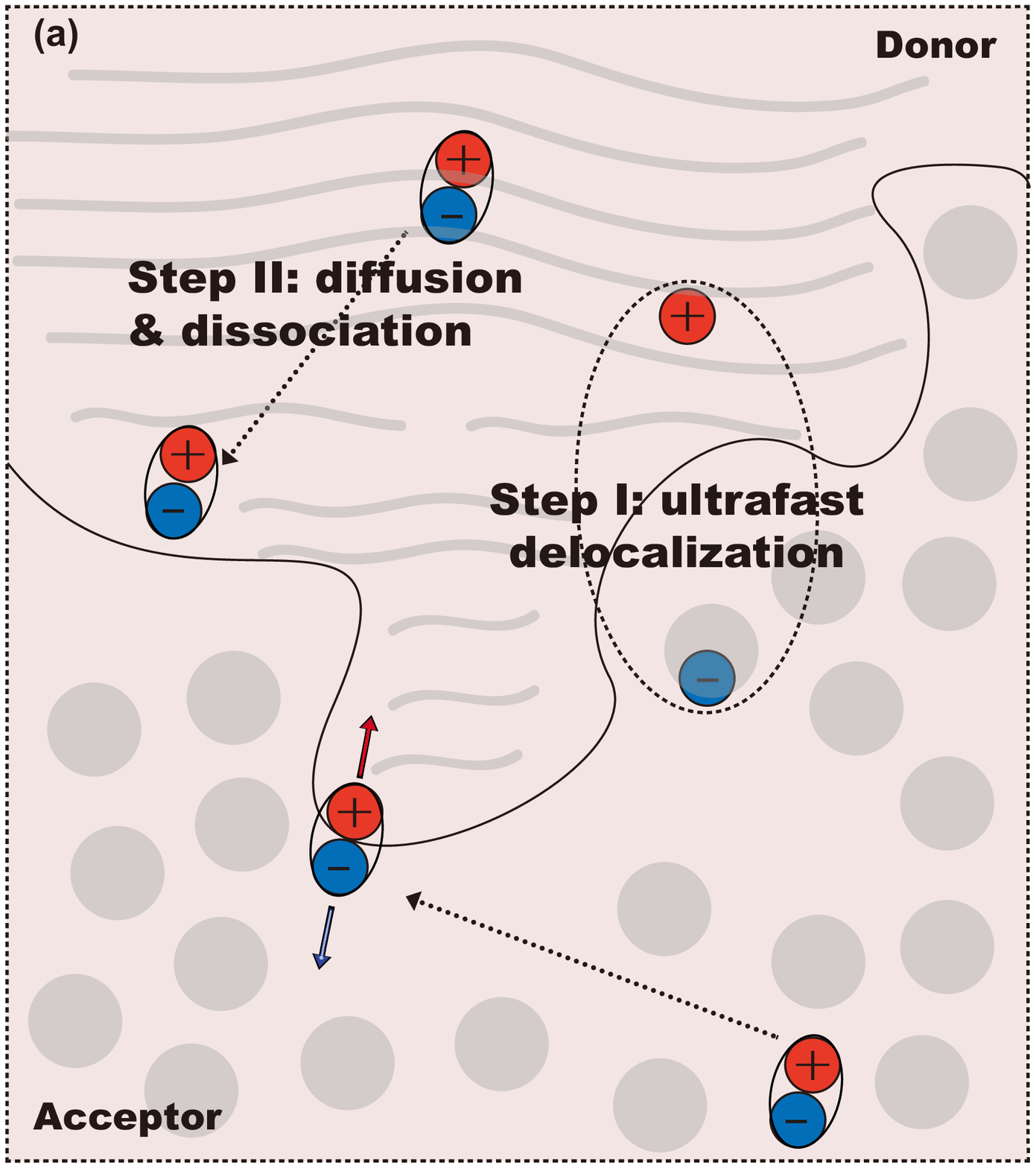}
\includegraphics[width=6cm]{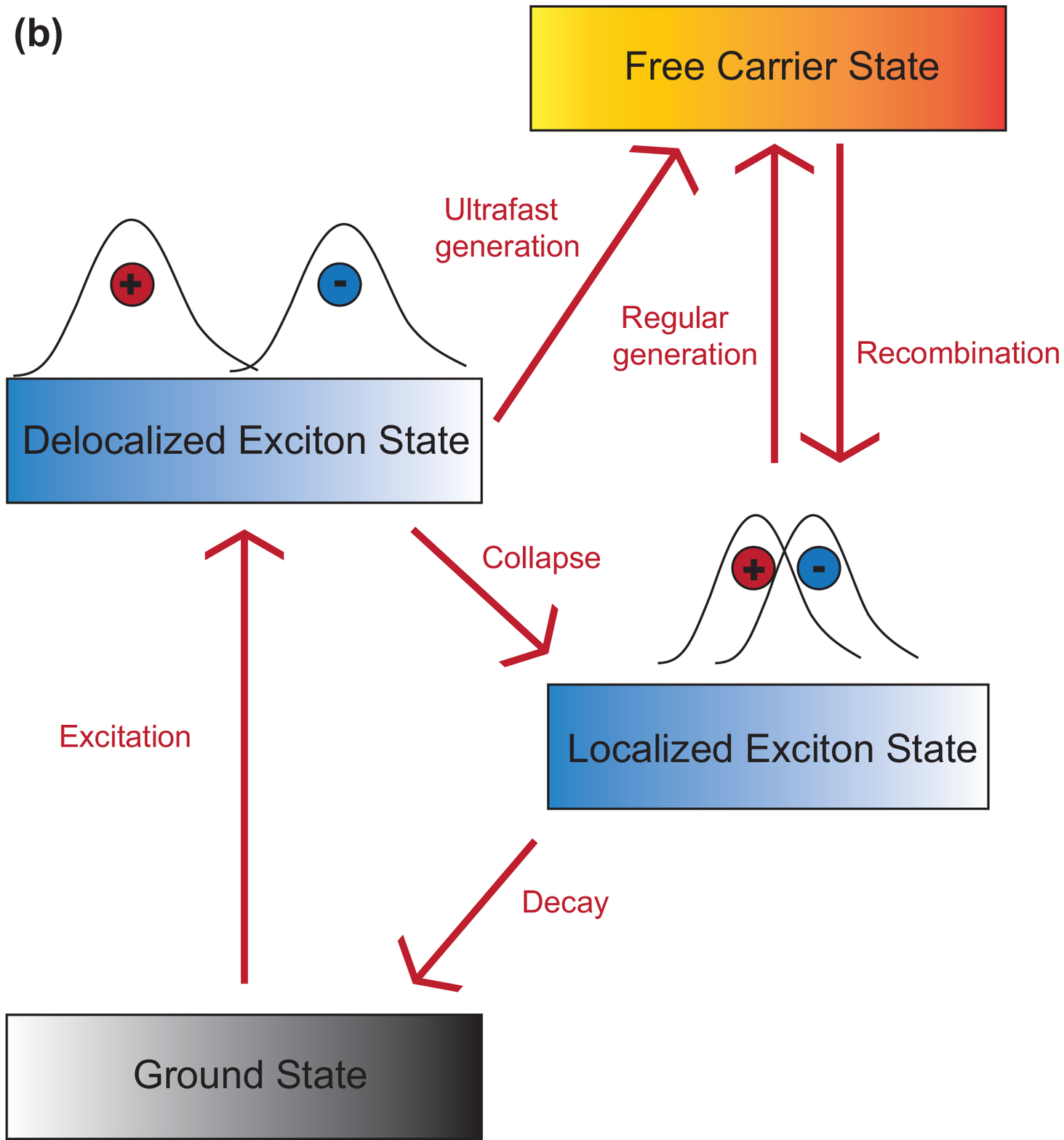}
\caption{\label{fig:1}(a) Charge-generation process in BHJ polymer materials. The ultrafast exciton delocalization (step I) and slow exciton diffusion and dissociation (step II) are illustrated. (b) Generation of free carriers in bulk-heterojunction OSCs.}
\end{figure}

Once the incident light is absorbed by the BHJ active material, an exciton will be created when an electron (e) is excited from the highest occupied molecular orbital (HOMO) to the lowest unoccupied molecular orbital (LUMO) of a molecule. Meanwhile, a hole (h) is created in the HOMO of the molecule. In organic materials, the spatial separation between the excited e and h is relatively small as compared to that in inorganic materials, and thus the excitons are referred to Frenkel excitons \cite{Excitonbook2}. In addition, Coulomb forces between the electrons and holes are strong. Therefore, the excitons are hard to be dissociated without excess energy. As described in the processes step II in Figure \ref{fig:1}(a), after being photogenerated, the excitons will diffuse from the created locations to the donor/acceptor interfaces \cite{Menke3,Feron4}. After overcoming the Coulomb attraction at the interfaces, the bounded electron-hole pairs will be dissociated, and the free carriers (electrons in the acceptor phase and holes in the donor phase) will be collected by electrodes. Regarding the excitons in the diffusion and dissociation processes, only singlet excitons are considered. Because triplet excitons have no contribution in the carrier generation process due to the difficulty in their dissociation. \cite{Menke3}

However, in contrast to the traditional view of charge separation from the exciton diffusion and dissociation, the existence of exciton delocalization has been confirmed by several groups recently \cite{Zhong5,Askat,Kaake6,Kaake7,Kaake8,Kaake9,Bakulin9,HeegerAM,Gelinas10}. Through the quantum mechanics/molecular simulation and experiments, researchers found an ultrafast charge transfer (generation) process at the donor-acceptor interfaces in a very short timescale ($<$ 100 fs) after the light absorption, as depicted in Figure \ref{fig:1}(a). As described in step I in Figure \ref{fig:1}(a), the excited exciton states have a large spatial delocalization, which means a large exciton radius. Thus the weak interaction between electrons and holes would facilitate the direct and ultrafast generation of free carriers. On the other hand, the experimental results showed that a part of the delocalization states will simultaneously collapse to the localized exciton states. After that, the localized (Frenkel) excitons continue to diffuse toward the BHJ interfaces, form the charge transfer states or polaron pairs, and finally dissociate to free carriers. The total processes of carriers generation are displayed in Figure \ref{fig:1}(b).

The timescale and ratio of the delocalization-assisted generation are determined by the donor-acceptor combinations intrinsically, and can also be adjusted by the fullerene aggregation \cite{FriendNew,Askat}. Additionally, the delocalization excitons also exist in non-fullerene polymer blended systems \cite{Bakulin9}. Those polymers and organic materials supporting the exciton delocalization have a great potential to be the next-generation high-performance photovoltaics.

After understanding the charge separation and generation processes, a systematic numerical model is established to fully capture the electrical responses of OSCs, involving exciton delocalization, exciton diffusion, dissociation and decay, as well as transport, recombination, and collection of free carriers. Particularly, the role of exciton delocalization in the performance of BHJ OSCs is investigated in detail. Regarding the device model developed, it does not resolve the microscopic delocalization process in an ultrafast time scale and ignores some intermediate stages, such as charge transfer states or polaron pairs. However, we will show that the exciton delocalization incorporated drift-diffusion model could well describe electrical responses of polymer solar cells. In our model, excitons are classified into two categories named localized excitons and delocalized excitions. The former satisfies the exciton diffusion-dissociation equation [See Eq.~\eqref{exciton}]; The latter will be directly inserted into the traditional drift-diffusion equation [See Eq.~\eqref{np}] as the generation rate of free carriers.

\section{Model}

\begin{figure}
  \includegraphics[width=3.0in]{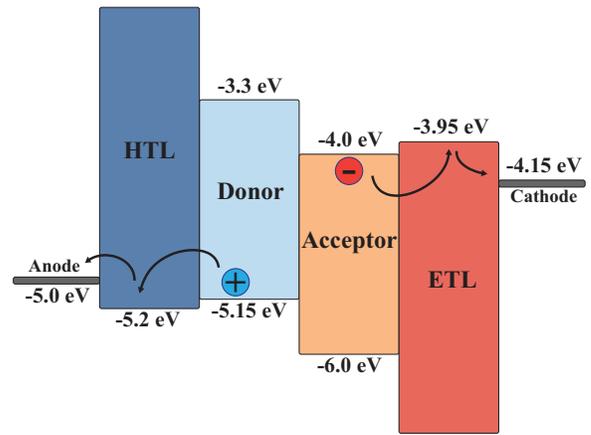}
  \caption{\label{fig:structure}Schematic illustration of the energy band diagram of the polymer solar cell to be modeled\cite{Caoyong}.}
\end{figure}

The OSC we studied has a structure of ITO anode/hole transport layer (HTL)/active layer (donor and acceptor)/electron transport layer (ETL)/Al cathorde. The active material is an admixture of high-performance organic material $\mathrm{PTB7}$, as the donor, and PC$_{70}$BM, as the acceptor. The schematic illustration of the energy band diagram is shown in Figure \ref{fig:structure}. The energy difference between the LUMO of the PCBM acceptor and the HOMO of the polymer donor is the band gap ($E_g$). The generation rates of localized and delocalized exciton states are assumed to be spatially homogeneous. Due to a very thin thickness of active layer in our study (110 nm), the unmatched incorporation of an exponential dependence of the generation rate on distance does not give rise to serious inconsistencies \cite{Koster11}. The free carriers are generated from two different physical processes. They are direct generation through the delocalization and indirect generation through the diffusion and dissociation, and the two processes will be treated separately. For the delocalized excitons the charge transfer in an ultrafast process, they do not involve the long timescale activity like the drift-diffusion of free carriers and diffusion-dissociation of excitons. Hence, the delocalization event could be simulated by directly introducing the exciton delocalization ratio into the carrier drift-diffusion and exciton diffusion-dissociation equations.

\begin{table}
\caption{\label{tab:table1}The model parameters used in the simulation.}.

\begin{ruledtabular}
\begin{tabular}{ccc}
Parameters          &Symbol         &Numerical{} Values\\
\hline
Bandgap                     &$E_{g}$                 &$1.15$ eV\\
Electron mobility\cite{Caoyong}           &$\mu_n$                 &$5.8\times10^{-7}\,\,\mathrm{m^2/Vs}$\\
Hole mobility\cite{Caoyong}                &$\mu_p$                 &$1.7\times10^{-7}\,\, \mathrm{m^2/Vs}$\\
Effective density of state	&$N_c$ ($N_v$)	         &$2.5\times10^{25}\,\, \mathrm{m^{-3}}$\\
Generation rate\cite{YuLuping1,GR}	            &$G$	                 &$1.08\times10^{28}\,\, \mathrm{m^{-3}}$\\
Exciton mobility\cite{lifetime}	        &$\mu_x$	             &$6.98\times10^{-6}\,\,\mathrm{m^2/Vs}$\\
Average dielectric constant	&$\varepsilon$	         &$3.9\varepsilon_0\,\,\mathrm{F/m}$\\
Exciton radius\cite{PTB7exciton}	            &$a$	                 &$3.5$ nm\\
Exciton lifetime\cite{lifetime}	        &$\tau_f$	             &$400\,\,p s$
\end{tabular}
\end{ruledtabular}
\end{table}

\subsection{Drift-diffusion of free carriers incorporating exciton delocalization}
The basic governing equation set for free carriers including Poisson equation, drift-diffusion equation, and current continuity equations read\cite{Koster11,Sha,NMYoung,Gummel,Mock}

\begin{equation}\label{phi}
\nabla  \cdot (\varepsilon \nabla \varphi ) =  - q(p - n)\\
\end{equation}

\begin{subequations}\label{np}
\begin{equation}\label{n}
\frac{{\partial n}}{{\partial t}} = {\eta _{d}}G +\frac{1}{q}\nabla  \cdot (q{\mu _n}n{E_n} + q{D_n}\nabla n) + {k_d}{X_l} - R(n,p)
\end{equation}

\begin{equation}\label{p}
\frac{{\partial p}}{{\partial t}} = {\eta _{d}}G - \frac{1}{q}\nabla  \cdot (q{\mu _p}p{E_p} - q{D_p}\nabla p) + {k_d}{X_l} - R(n,p)
\end{equation}
\end{subequations}

\noindent where $q$ is the elementary charge and $\varepsilon$ is the dielectric constant. $\mu_{n,p}$ are the mobility of electrons and holes, and $D_{n,p}$ are the corresponding diffusion coefficients, which obey the Einstein relation:

\begin{equation}\label{Ein}
{D_{n,p}} = \frac{{{k_B}T}}{q}{\mu _{n,p}}\\
\end{equation}

\noindent where $k_B$ is the Boltzmann constant and $T$ is the absolute temperature. $E_{n,p}$ are the internal electrostatic fields for electrons and holes, $\eta_{d}$ is the exciton delocalization ratio, and $G$ is the exciton generation rate at the active layer. $X_l$  and $k_d$ are the density and dissociation rate of the localized singlet excitons, respectively.

$R(n,p)$ in Eq.~\eqref{np} is the recombination rate of free carriers, which is described to be the bimolecular process, i.e.

\begin{equation}\label{recom}
R = {r_R} \cdot \gamma (np - n_{i}^2)
\end{equation}
where $n_{i}$ is the intrinsic carrier density of the active material, and $\gamma$ is the recombination strength by the Langevin \cite{Langevin12},

\begin{equation}\label{Recomfactor}
\gamma  = \frac{q}{{\left\langle \varepsilon  \right\rangle }}\left\langle \mu  \right\rangle
\end{equation}

\noindent where $\left\langle \varepsilon  \right\rangle$ is the spatially averaged dielectric constant and $\left\langle \mu  \right\rangle$ is the spatially averaged summation of hole and electron mobility. The factor $r_R$ in Eq.~\eqref{recom} is introduced to coordinate the recombination rate due to the fact that the recombination loss in the experiments is much lower than that predicted by the Langevin theory\cite{Tachiya,Burke13}. Here we set $r_R$= 0.01 in the simulation, which is a proper approximation for PTB7:PC$_{70}$BM.

\subsection{Diffusion-dissociation of localized excitons}
The basic governing equation for localized excitons is given by:

\begin{equation}\label{exciton}
\frac{{\partial {X_l}}}{{\partial t}} = (1 - {\eta _{d}})G + \nabla  \cdot ({D_X}\nabla {X_l}) - {k_d}{X_l} -\frac{X_l}{\tau_f}+ {\eta _s}R(n,p)
\end{equation}

\noindent where $1-\eta_{d}$  is the exciton localization ratio. $D_X$ is the exciton diffusion coefficient related to the exciton mobility $\mu_X$  by the Einstein relation, and $\tau_f$ is the lifetime of a singlet exciton. By the F\"orster energy transfer in the polymer chains \cite{Bakulin9}, the localized excitons diffuse, until reaching the interfaces of the n-p material. They will dissociate, which is expressed by the factor  $k_dX_l$, or will decay to the ground states, expressed by $X_l$/$\tau_f$ . Moreover, the free carriers also recombine into the localized excitons, and only singlet excitons are considered with $\eta_s$ of 1/4\cite{Excitonbook2,singlet,Menke3}.

The theory of germinate recombination, originally discussed by Onsager\cite{Onsager14,Onsager15} and refined by Braun\cite{Braun16}, proposed that the probability of exciton dissociation depends on the temperature, distance and field,

\begin{equation}\label{probability}
p(x,F,T) = \frac{{{k_d}(x,F,T)}}{{{k_d}(x,F,T) + {k_f}(T)}}
\end{equation}

\noindent where $T$ is the temperature, $x$ is the distance between the bound charges of the exciton, $F$ is the internal electrostatic field, and $k_f$ is the exciton decay rate, which equals to $1/\tau_f$, and is attributed to the germinate recombination. Braun derives the following expression for $k_d$,

\begin{equation}\label{kd}
{k_d}(x,F,T) = \frac{{3\gamma }}{{4\pi {x^3}}}{e^{ - {{{E_B}} \mathord{\left/
 {\vphantom {{{E_B}} {{k_B}T}}} \right.
 \kern-\nulldelimiterspace} {{k_B}T}}}}{J_1}(2\sqrt { - 2b} )/\sqrt { - 2b}
\end{equation}

\noindent where $E_B=q^2/4\pi\varepsilon x$ is the exciton binding energy, $J_1$ is the first-order Bessel function, and the field parameter $b= q^3 F/(8\pi\varepsilon k_B^2 T^2)$.  The disordered polymer BHJ system can be simulated by different ways. On one hand, the random distribution is modeled by Monte Carlo method \cite{Meng17,Stefan18}. On the other hand, the structure also makes itself appropriate to treat the charge-separation distance as a Gaussian distribution \cite{VD19}, where the overall exciton dissociation probability comes from the integral average over all the charge-separation distances,
\begin{equation}\label{overallP}
P(F,T) = \frac{4}{\sqrt\pi {a^3}}\int_0^\infty p(x,F,T){x^2}e^{-(x/a)^2}dx
\end{equation}

\noindent where $a$ is the charge-separation distance at the maximum probability of Gaussian function, which is also regarded as the exciton radius. Then we got the changed exciton dissociation rate,

\begin{equation}\label{rate}
{k_d}(x,F,T) = \frac{{P(F,T)}}{{1 - P(F,T)}}{k_f}(T)
\end{equation}

\subsection{Boundary conditions}
In order to numerically solve the coupled nonlinear system including equations~\eqref{phi}, \eqref{np} and \eqref{exciton}, the boundary conditions shall be defined. For convenience, we set $x=0$ as the bottom cathode of the solar cells, where $x$ denotes the position, and set $x=L$ as the top anode, where $L$ is the thickness of the device. Both electrodes are the Schottky contacts satisfying the following boundary conditions

\begin{equation}
n(0) = {N_C}{\exp{ \left(- \frac{{{U_{BN}}}}{{{k_B}T}}\right)}}
\end{equation}
\begin{equation}
n(L) = {N_C}{\exp{\left[\frac{{({U_{BP}} - {E_{g}})}}{{{k_B}T}}\right]}}
\end{equation}

\begin{equation}
p(0) = {N_V}{\exp{\left[\frac{{({U_{BN}} - {E_{g}})}}{{{k_B}T}}\right]}}
\end{equation}
\begin{equation}
p(L) = {N_V}{\exp{ \left(- \frac{{{U_{BP}}}}{{{k_B}T}}\right)}}
\end{equation}

\noindent where $N_C$ and $N_V$ are the effective density of states of conduction band and valence band of the carrier transport layers (CTLs), and $U_{BN}$ and $U_{BP}$ are the injection barriers at the electrodes. $E_{g}$ is the bandgap. The boundary conditions for the potential should be

\begin{equation}
\psi (0) =  - {W_C}, \quad \psi (L) = {V_{a}} - {W_A}
\end{equation}

\noindent where $W_C$ and $W_A$ are the work functions of cathode and anode, and $V_{a}$ is the applied voltage. The excitons quench at the interfaces between CTLs and active layer ($x=d_1$, $x=d_2$) satisfying

\begin{equation}
{X_l}(d_1) = {X_l}(d_2) = 0
\end{equation}

\section{Simulation results and discussions }
\begin{figure}
\includegraphics[width=3.0in]{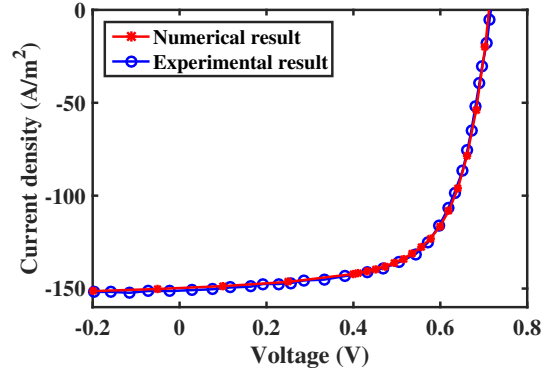}
\caption{\label{fig:PTB7}Theoretical (stars) and experimental (circles) current density-voltage characteristics of PTB7:PC$_{70}$BM device. The experimental result is extracted from the literature [29].}
\end{figure}

All the simulation parameters are listed in Table \ref{tab:table1} and the injection barriers $U_{BN}$ and $U_{BP}$ are 0.2 eV. The coupled nonlinear partial differential equations~\eqref{phi}, \eqref{np} and \eqref{exciton} are solved self-consistently by the Scharfetter-Gummel method in the space domain and by the semi-implicit scheme in the time domain \cite{Gummel}. The detailed computational methods are shown in the Supplementary Material. From this method we can obtain both the transient and steady solutions, but in this study we focus on the steady result of the OSC. The numerical results based on this model match well with the experiments\cite{PTB7}, under the condition of 75\% delocalization ratio. The ratio of 70\% - 80\% is demonstrated to be an appropriate approximation for high-performance organic materials\cite{Kaake6,Kaake9}. The current-voltage curves of numerical and experimental results are shown in Figure \ref{fig:PTB7}, which confirms the validity of our model.

\subsection{Exciton delocalization ratio dependent electrical properties}

\begin{figure}
\includegraphics[width=3.0in]{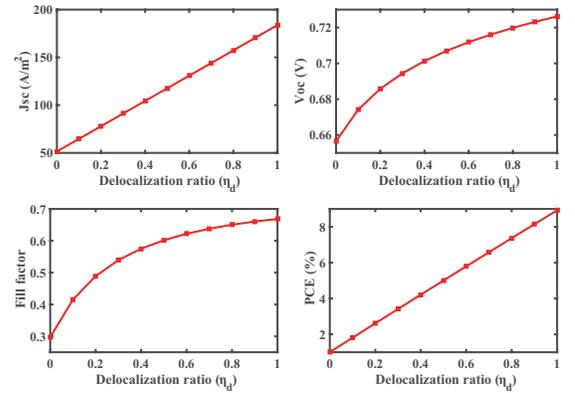}
\caption{\label{fig:compare}Short-circuit current (a), open-circuit voltage (b), fill factor (c), and power conversion efficiency (d) of the simulated device.}
\end{figure}
In Figures \ref{fig:compare}(a)-(d), the electrical responses of the OSC are calculated as a function of the delocalization ratio. With the increase of the delocalization ratio, the short-circuit current ($J_{sc}$), open-circuit voltage ($V_{oc}$), Fill Factor (FF) and power conversion efficiency (PCE) are all increased. The most significant change is the short-circuit current that is improved over 3 times from 51.4 $\mathrm {A/m^2}$ to 183.8 $\mathrm {A/m^2}$, almost linearly with the delocalization ratio. The PCE will be improve from 1.0\% to 8.9\%, while the PCE of well-made PTB7-PC$_{70}$BM solar cells in reality is near 7\%. The main energy loss in OSCs occurs at the exciton diffusion-dissociation process, and is mainly caused by a finite exciton dissociation probability and unseparated excitons that will decay to their ground states producing heat. The reduction of the localized excitons will reduce the thermodynamic loss of OSCs and accordingly will boost the free carrier generation, which is essential to high-performance OSCs.

\begin{figure}
\includegraphics[width=3.0in]{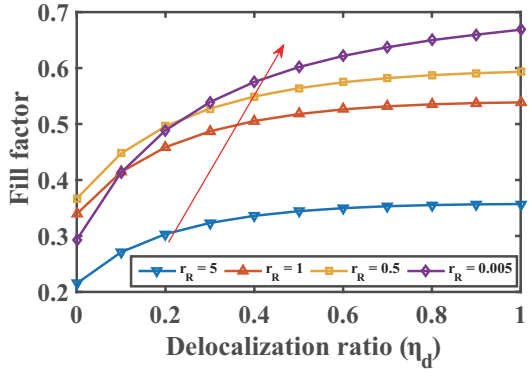}
\caption{\label{fig:FFchange}Recombination rate dependent fill factor saturation. A large $r_R$ means a large bulk recombination rate.}
\end{figure}

The enormous increase of $J_{sc}$ plays the dominant role in the improved PCE, and thus the PCE keeps a linear trend. Interestingly, in the Figure \ref{fig:compare}(c), FF is saturated with the rise of the delocalization ratio. Fill factor represents the balance between carrier extraction and recombination. On one hand, the higher the delocalization ratio is, the more free carriers are generated, and less localized excitons will be formed. On the other hand, more free carriers generated will induce a higher bimolecular recombination and therefore, more localized excitons decay to their ground states. Consequently, the increasing rate of FF will slow down, and the FF will reach a saturation point. This interpretation can be supported by the simulation result shown in Figure \ref{fig:FFchange}. When reducing the recombination rate in the device by decreasing the factor $r_R$ from 5 to 0.005 in Eq.~\eqref{recom}, the overall recombination in the device will decrease, thus the saturation will be reached at a high fill factor.

\subsection{Device responses under $J_{sc}$ and $V_{oc}$ conditions}

\begin{figure}
\includegraphics[width=3.0in]{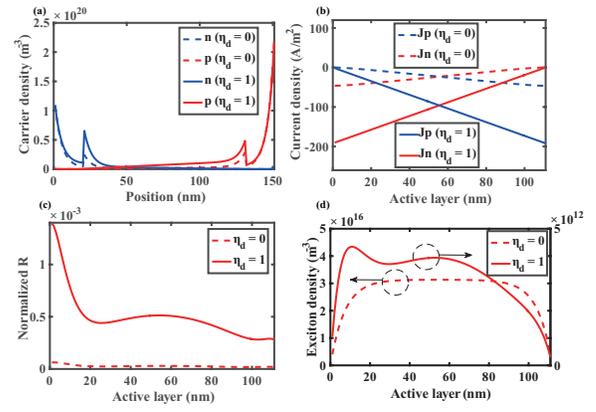}
\caption{\label{fig:jsc} Carriers density (a), current density(b), recombination rate (normalized to generation) (c) and exciton density (d) distributions in the device (a) or acitve layer (b,c,d) under the short circuit condition with the delocalization ratios of 0\% and 100\%.}
\end{figure}

\begin{figure}
\includegraphics[width=3.0in]{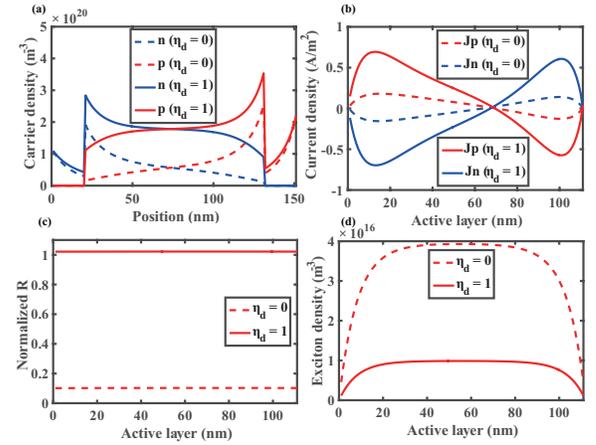}
\caption{\label{fig:voc}Carriers density (a), current density(b), recombination rate (normalized to generation) (c) and exciton density (d) distributions in the device (a) or acitve layer (b,c,d) under the open circuit condition with the delocalization ratios of 0\% and 100\%.}
\end{figure}
In order to further explain the effect of delocalization, the calculated electron/hole density, current density, recombination rate (normalized to generation) and exciton density distribution in the device region or acitve layer under the conditions of the short-circuit (SC) ($V=0$) and open-circuit (OC) ($V=Voc$) are depicted in Figures \ref{fig:jsc} and \ref{fig:voc}, respectively. Due to the extraction barriers at the interfaces between the active layer and CTLs, and also the Schottky injection barriers at the electrodes, the free carriers will accumulate there. When increasing the external applied voltage, the reduced potential makes free carriers accumlate in a broaden region. Moreover, the peaks of the charge densities at the two electrodes are unbalanced due to the unbalanced mobility of electrons and holes. (See Figures. \ref{fig:jsc}(a) and \ref{fig:voc}(a)). Under the SC state, the large internal electrostatic field causes a linear drift current density distribution at the active layer region, while under the OC state, due to the zero internal electrostatic field, the electron and hole diffusion currents are quite small and cancel with each other at every point, and thus the tole current is zero, as shown in Figures. \ref{fig:jsc}(b) and \ref{fig:voc}(b). The carriers accumulation induces a larger recombination rate at the interfaces between the active layer and CLTs [See Figure \ref{fig:jsc}(c)]. Additionally, the CTLs help to suppress the recombination around the electrodes, therefore the normalized bulk recombination rate in the complete delocalization case can reach the unit everywhere to cancel the generation at the OC case as shown in Figure \ref{fig:voc}(c). The normalized bulk recombination rate in the localization case is much lower than the unit. It is because that the big amount of excitons have large germinate recombination losses.

In all the cases, the large exciton delocalization ratio induces the high density of free carriers and thus the boosted current, even though with the higher recombination rate. According to Figure \ref{fig:1}, the localized excitons are generated by a direct collapse of delocalization excitons and by an indirect bimolecular recombination, with respect to the terms of $(1-\eta_{d})G$ and $\eta_s R$, respectively. When increasing the external applied voltage, the first part of the localized excitons is constant, while the second part will be increased by the recombination. Meanwhile, according to Eq.~\eqref{kd}, the field-dependent exciton dissociation probability becomes extremely lower at the OC case where the internal electrostatic field is almost zero. Thus, from the SC to the OC cases, the increased recombination as well as the decreased dissociation leads to the enlarged exciton density. At the SC state, the exciton density for the complete delocalization case is 4 orders smaller than that of localization case. However, at the OC state, it is only 4 times smaller than that of the localization case, as illustrated in Figures \ref{fig:jsc}(d) and \ref{fig:voc}(d). Because the excitons have the low probability of being separated into free carriers.
\section*{Supplementary Material}
    See supplementary material for the detailed computational method for the exciton delocalization incorporated drift-diffusion model.
\section{Conclusion}
In conclusion, we studied the role of exciton delocalization in the performance of OSCs, and developed a systematic exciton delocalization-diffusion-dissociation and carrier drift-diffusion-recombination model. We found $J_{sc}$, $V_{oc}$, FF and PCE are simultaneously improved when increasing the exciton delocalization ratio. More importantly, the saturations of FF were also investigated and understood. The simulation results suggest that the enhancement of the delocalization ratio by optimizing the BHJ structure is essential to the performance of OSCs. Our work is very helpful to high-efficiency OSCs.

\begin{acknowledgments}
The project was supported by The National Natural Science Foundation of China (No. 61201122). It was also supported by the University Grant Council of the University of Hong Kong (Grant No. 201311159056), the General Research Fund (Grant Nos. HKU711813 and HKU711612E), the Collaborative Research Fund (Grant No. C7045-14E) and RGC-NSFC Grant ($\mathrm{N\_HKU709/12}$) from the Research Grants Council of Hong Kong Special Administrative Region, China, and Grant No. CAS14601 from CAS-Croucher Funding Scheme for Joint Laboratories. This project was supported in part by Hong Kong UGC Special Equipment Grant (SEG HKU09).
\end{acknowledgments}

\end{document}


\author{Zi Shuai Wang, Wei E. I. Sha\thanks{wsha@eee.hku.hk}, and Wallace C. H. Choy\thanks{chchoy@eee.hku.hk}}
\maketitle
\begin{center}
\textbf{\LARGE Supporting Information}
\end{center}

This supplementary material offers the computational method used in solving the exciton delocalization incorporated drift-diffusion model.

In this model, we used the following equations:
\setcounter{equation}{0}
\renewcommand{\theequation}{S\arabic{equation}}
\begin{equation}\label{phi}
\nabla  \cdot (\varepsilon \nabla \varphi ) =  - q(p - n)\\
\end{equation}

\begin{equation}\label{n}
\frac{{\partial n}}{{\partial t}} = {\eta _{d}}G +\frac{1}{q}\nabla  \cdot (q{\mu _n}n{E_n} + q{D_n}\nabla n) + {k_d}{X_l} - R(n,p)
\end{equation}
\begin{equation}\label{p}
\frac{{\partial p}}{{\partial t}} = {\eta _{d}}G - \frac{1}{q}\nabla  \cdot (q{\mu _p}p{E_p} - q{D_p}\nabla p) + {k_d}{X_l} - R(n,p)
\end{equation}

\begin{equation}\label{exciton}
\frac{{\partial {X_l}}}{{\partial t}} = (1 - {\eta _{d}})G + \nabla  \cdot ({D_X}\nabla {X_l}) - {k_d}{X_l} -\frac{X_l}{\tau_f}+ {\eta _s}R(n,p)
\end{equation}
where the physical explanations of all the parameters are described in the manuscript. Here, we will present the numerical method to discretize the above partial differential equation set. We used the Scharfetter-Gummel scheme in the spatial domain and the semi-implicit strategy in the temporal domain~\cite{Book,Sha}. The one-dimensional discretized forms of Eqs. \eqref{phi} - \eqref{exciton} are respectively given by:

\begin{equation}\label{dphi}
\begin{aligned}
\frac{1}{\Delta y^2}\varepsilon_{i+1/2}\varphi_{i+1}^{m+1}-\left[\frac{2}{\Delta y^2}\left(\varepsilon_{i+1/2}+\varepsilon_{i-1/2}\right)+\frac{n_i^m+p_i^m}{U_t}\right]\varphi_{i}^{m+1}+ \\\frac{1}{\Delta y^2}\varepsilon_{i-1/2}\varphi_{i-1}^{m+1}=-q\left(p_i^m-n_i^m\right)-\frac{n_i^m+p_i^m}{U_t}{\varphi}_i^m
\end{aligned}
\end{equation}
\begin{equation}\label{dn}
\begin{aligned}
\left[\frac{1}{\Delta t}+D_{n,i+1/2}B\left(\frac{\varphi_{n,i}^{m+1}-\varphi_{n,i+1}^{m+1}}{U_t}\right)+D_{n,i-1/2}B\left(\frac{\varphi_{n,i}^{m+1}-\varphi_{n,i-1}^{m+1}}{U_t}\right)\right] n_i^{m+1}\\
-D_{n,i+1/2}B\left(\frac{\varphi_{n,i+1}^{m+1}-\varphi_{n,i}^{m+1}}{U_t}\right)n_{i+1}^{m+1}-D_{n,i-1/2}B\left(\frac{\varphi_{n,i-1}^{m+1}-\varphi_{n,i}^{m+1}}{U_t}\right)n_{i-1}^{m+1}\\
=\frac{1}{\Delta t}n_i^{m}+\eta_dG-R\left(n_i^m,p_i^m\right)+k_{d,i}X_{l,i}^m
\end{aligned}
\end{equation}
\begin{equation}\label{dp}
\begin{aligned}
\left[\frac{1}{\Delta t}+D_{p,i+1/2}B\left(-\frac{\varphi_{p,i}^{m+1}-\varphi_{p,i+1}^{m+1}}{U_t}\right)+D_{p,i-1/2}B\left(-\frac{\varphi_{p,i}^{m+1}-\varphi_{p,i-1}^{m+1}}{U_t}\right)\right]  p_i^{m+1}\\
-D_{p,i+1/2}B\left(-\frac{\varphi_{p,i+1}^{m+1}-\varphi_{p,i}^{m+1}}{U_t}\right)p_{i+1}^{m+1}-D_{p,i-1/2}B\left(-\frac{\varphi_{p,i-1}^{m+1}-\varphi_{p,i}^{m+1}}{U_t}\right)p_{i-1}^{m+1}\\
=\frac{1}{\Delta t}p_i^{m}+\eta_dG-R\left(n_i^m,p_i^m\right)+k_{d,i}X_{l,i}^m
\end{aligned}
\end{equation}
\begin{equation}\label{dx}
\begin{aligned}
\left[\frac{1}{\Delta t}+\frac{D_{X,i+1/2}+D_{X,i-1/2}}{\Delta y^2}+k_{d,i}+\frac{1}{\tau_f}\right] X_{l,i}^{m+1}-\frac{D_{X,i+1/2}}{\Delta y^2}X_{l,i+1}^{m+1}-\\ \frac{D_{X,i-1/2}}{\Delta y^2}X_{l,i-1}^{m+1}
=\frac{1}{\Delta t}X_{l,i}^m+\left(1-\eta_d\right)G+\eta_sR\left(n_i^m,p_i^m\right)
\end{aligned}
\end{equation}
where the superscript $m$ and subscript $i$ indicate the discretized temporal and spatial steps, respectively. Meanwhile, in these equations, $U_t=\frac{k_BT}{q}$, $B(x)=\frac{x}{e^x-1}$, and the electron potential and hole potential are of the forms
\begin{equation}\label{npotential}
  \varphi_n=\varphi+\frac{X_i}{q}+\frac{k_BT}{q}\ln(N_c)
\end{equation}

\begin{equation}\label{ppotential}
  \varphi_p=\varphi+\frac{X_i}{q}+\frac{E_g}{q}-\frac{k_BT}{q}\ln(N_v)
\end{equation}

\noindent where $X_i$ is the electron affinity, $E_g$ is the bandgap, and $N_{c,v}$ indicates the effective density of states.

For every iteration in the time domain, the Poisson's equation will be solved first, and then the exciton diffusion-dissociation equation, electron and hole drift-diffusion equations will be solved in turn. The iteration will continue until a steady solution is obtained. The convergence condition for the steady solution calculation is given by $\left|\frac{J^{m+1}-J^m}{J^{m}}\right|<10^{-4}$.
\setcounter{figure}{0}
\renewcommand{\thefigure}{S\arabic{figure}}
\begin{figure}
  \centering
  \includegraphics[width=3 in]{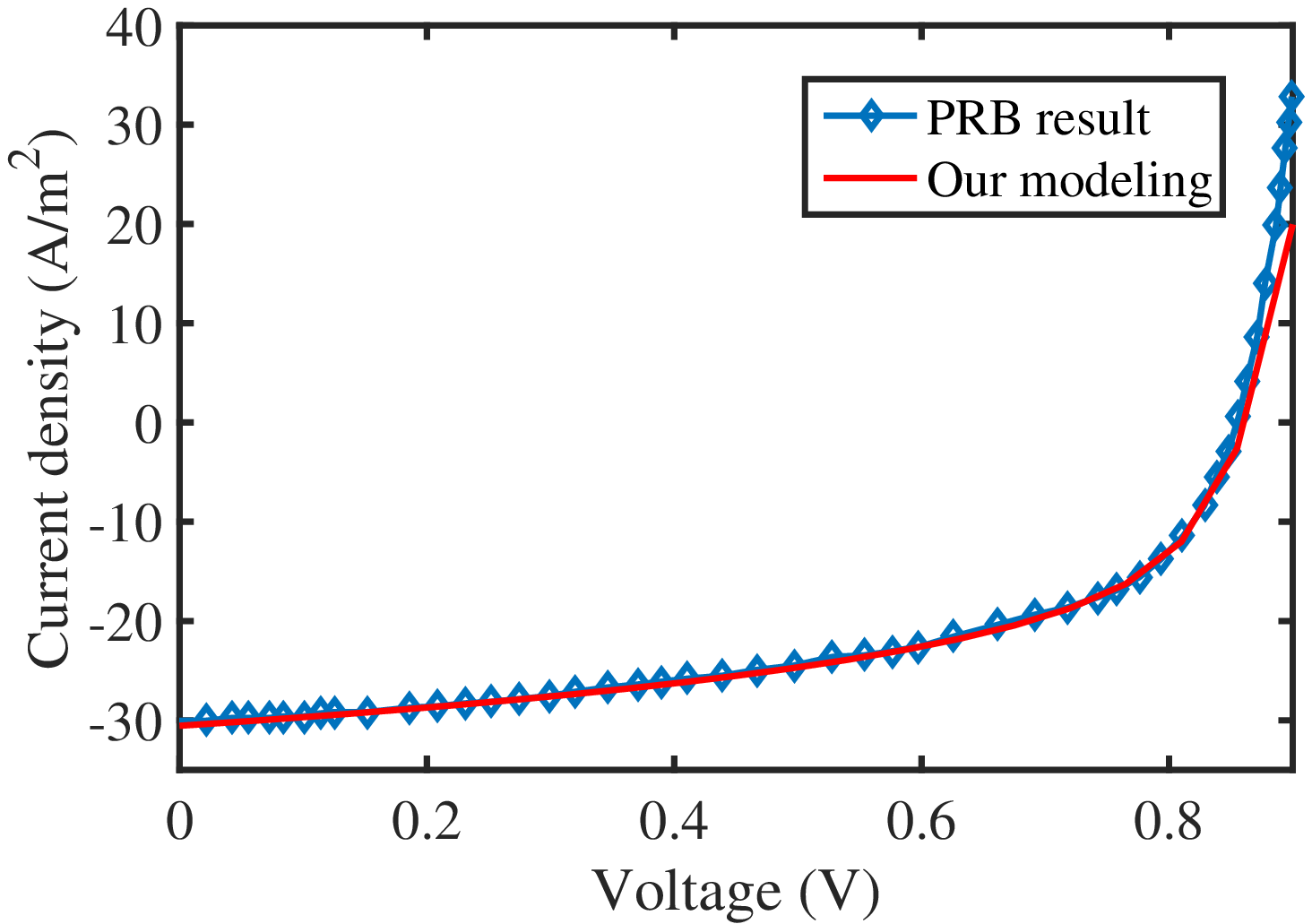}
  \caption{The current density-voltage characteristics of a polymer bulk-heterojunction solar cell. The simulation parameter can be found in [Phys. Rev. B 72: 085205, 2005.]}\label{JV}
\end{figure}

As shown in Fig.~\ref{JV}, using this numerical method, our theoretical result has a good agreement with the result extracted from the literature [L. J. A. Koster et al., Phys. Rev. B 72: 085205, 2005] by using the same exciton model in the literature. Thus, the numerical method adopted is convincing to obtain the steady solution of polymer solar cells including current density-voltage characteristics. Additionally, by setting a time-dependent generation rate, the numerical method also can be utilized to simulate transient photovoltage (TPV) or transient photocurrent (TPC) of polymer solar cells, where the excitation is not sunlight but a short laser pulse.